\documentclass[12pt]{iopart}

\usepackage[caption=false]{subfig}

\usepackage[T1]{fontenc}
\usepackage[utf8]{inputenc}
\usepackage[english]{babel}

\usepackage{graphicx}
\usepackage{enumitem}
\usepackage{textcomp}
\usepackage{gensymb}
\usepackage{siunitx}
\usepackage{isotope}
\usepackage{rotating}

\usepackage[sort]{natbib}
\expandafter\let\csname equation*\endcsname=\relax
\expandafter\let\csname endequation*\endcsname=\relax
\usepackage{amsmath}

\usepackage{amsfonts,amssymb,amscd}
\usepackage{xcolor, soul}
\usepackage{xspace}
\usepackage{microtype}

\usepackage{hyperref}
\usepackage{cleveref} 

\usepackage{fdsymbol}

\sethlcolor{yellow}

\hypersetup{colorlinks, linkcolor={red!50!black}, citecolor={blue!50!black}, urlcolor={blue!80!black}}
\microtypesetup{
	protrusion=alltext-nott,
	expansion=alltext-nott,
	final
}

\graphicspath{{figures/}}

\usepackage[textsize=tiny,disable]{todonotes}

\begin{document}
\title[GATE 10 - Part I]{GATE 10 Monte Carlo particle transport simulation - Part I: development and new features}

\author{
David Sarrut\textsuperscript{1}, 
Nicolas Arbor\textsuperscript{9},
Thomas Baudier\textsuperscript{1},
Julien Bert\textsuperscript{4},
Konstantinos Chatzipapas\textsuperscript{4,16},
Martina Favaretto\textsuperscript{5},
Hermann Fuchs\textsuperscript{10},
Loïc Grevillot\textsuperscript{5},
Hussein Harb\textsuperscript{4},
Gert Van Hoey\textsuperscript{17},
Maxime Jacquet\textsuperscript{1},
Sébastien Jan\textsuperscript{6},
Yihan Jia\textsuperscript{5},
George C. Kagadis\textsuperscript{12},
Han Gyu Kang\textsuperscript{3},
Paul Klever\textsuperscript{8,13},
Olga Kochebina\textsuperscript{6},
Wojciech Krzemien\textsuperscript{18},
Lydia Maigne\textsuperscript{7},
Philipp Mohr\textsuperscript{13},
Guneet Mummaneni\textsuperscript{2},
Valentina Paneta\textsuperscript{11},
Panagiotis Papadimitroulas\textsuperscript{11},
Alexis Pereda\textsuperscript{7},
Axel Rannou\textsuperscript{4},
Andreas F. Resch\textsuperscript{5},
Emilie Roncali\textsuperscript{2},
Maxime Toussaint\textsuperscript{14},
Carlotta Trigila\textsuperscript{2},
Charalampos Tsoumpas\textsuperscript{13},
Jing Zhang\textsuperscript{4},
Karl Ziemons\textsuperscript{8},
Nils Krah\textsuperscript{1,15}
}

\address{
\textsuperscript{1} Université de Lyon; CREATIS; CNRS UMR5220; Inserm U1294; INSA-Lyon; Université Lyon 1, Lyon, France.\\
\textsuperscript{2} University of California, Davis, Davis CA USA\\
\textsuperscript{3} National Institutes for Quantum Science and Technology (QST), 4-9-1 Anagawa, Inage-ku, Chiba, Japan\\
\textsuperscript{4} LaTIM, INSERM UMR1101, University of Brest, Brest, France\\
\textsuperscript{5} MedAustron Ion Therapy Center, Wiener Neustadt, Austria\\
\textsuperscript{6} Université Paris-Saclay, Inserm, CNRS, CEA, Laboratoire d’Imagerie Biomédicale Multimodale (BioMaps), Orsay, France\\
\textsuperscript{7} Université Clermont Auvergne, Laboratoire de Physique de Clermont Auvergne, CNRS, Clermont-Ferrand, France.\\
\textsuperscript{8} FH Aachen University of Applied Sciences, Germany\\
\textsuperscript{9} Université de Strasbourg, IPHC, CNRS, UMR7178, F-67037 Strasbourg, France.\\
\textsuperscript{10} Medical University of Vienna, Vienna, Austria\\
\textsuperscript{11} Bioemission Technology Solutions, BIOEMTECH, Athens, Greece\\
\textsuperscript{12} 3DMI Research Group, Department of Medical Physics, University of Patras, Rion, Greece\\
\textsuperscript{13} University of Groningen, University Medical Center Groningen, Groningen, Netherlands\\
\textsuperscript{14} Laboratoire CRCI2NA, INSERM, CNRS, Nantes Université, Nantes, France\\
\textsuperscript{15} now with Department of Research and Development, Holland Proton Therapy Centre Delft, Delft, The Netherlands\\
\textsuperscript{16} Department of Radiation Science and Technology, Technical University of Delft, Delft, The Netherlands\\
\textsuperscript{17} XEOS, Ghent, Belgium\\
\textsuperscript{18} High Energy Physics Division, National Centre for Nuclear Research, Andrzeja Soltana 7, Otwock, Swierk, PL-05-400, Poland\\
}

\newpage

\begin{abstract}

We present GATE version 10, a major evolution of the open-source Monte Carlo simulation application for medical physics, built on Geant4.
This release marks a transformative evolution, featuring a modern Python-based user interface, enhanced multithreading and multiprocessing capabilities, the ability to be embedded as a library within other software, and a streamlined framework for collaborative development.
In this Part 1 paper, we outline GATE's position among other Monte Carlo codes, the core principles driving this evolution, and the robust development cycle employed.
We also detail the new features and improvements.
Part 2 will detail the architectural innovations and technical challenges.
By combining an open, collaborative framework with cutting-edge features, such a Monte Carlo platform supports a wide range of academic and industrial research, solidifying its role as a critical tool for innovation in medical physics.

\end{abstract}



\section{Introduction}
\label{sec:intro}

For about 20 years, the GATE software, developed by the OpenGATE collaboration, has been utilized by hundreds of projects and thousands of users worldwide to model applications in nuclear medicine, radiotherapy, and radiology, as well as in other fields like space radiation. GATE has played a crucial role in numerous academic research projects and industrial developments, leading to hundreds of scientific publications since its inception. Built on Geant4~\citep{allison2016recent}, GATE has been described in several reference papers throughout its history~\citep{jan2004, jan2011, sarrut2014a, sarrut2021a}. GATE fully relies on Geant4 for its Monte Carlo engine and provides: 1) easy access to Geant4 functionalities, and 2) additional features (e.g., variance reduction techniques) specifically designed for medical physics. Through the years, GATE has evolved to the major version 9, and specifically to 9.4. More technical details about the version 9.x series are available in~\citep{sarrut2022}.

More recently, the collaboration has faced some challenges associated with maintaining a large codebase (around 300,000 lines of code) developed over time by numerous contributors, most of whom contributed during their PhD or postdoctoral research projects. The current GitHub repository lists more than 80 unique contributors, although this repository was only established around 2012, meaning early contributors are not fully accounted for. This diversity has fostered innovation and experimentation, but has also led to maintenance issues. Some parts of the code have been "abandoned", while others are duplicated. Additionally, the C++ language has evolved significantly over the past 20 years, introducing more efficient and convenient constructs such as smart pointers, lambda functions, or the \texttt{auto} keyword, which improve code robustness and maintainability.

Taking into account the core principles of GATE (community-driven, open-source, and focused on medical physics), we embarked on a project to propose a new way of executing Monte Carlo simulations in medical physics, including cross-platform capability. The goal of GATE 10 was to provide a simple yet flexible Python-based interface through which users can set up and run Geant4 simulations. Internally, GATE 10 aims to offer developers and contributors a structured framework that simplifies the implementation and maintenance of new features. Although this radical change inevitably breaks immediate compatibility with previous versions, it preserves the core philosophy of simulation description.

In this article (Part 1), we introduce the guiding principles, goals, and technical foundations of the GATE 10 project. We also describe the new functionalities and their potential impact on the medical physics community. The companion paper~\citep{krah2025} provides a more detailed exploration of the architectural innovations and technical challenges faced during the development of GATE 10.


\section{General considerations about the development of GATE~10}

\subsection{User language: Python}

Building Monte Carlo simulation applications with a general-purpose toolkit like Geant4 is a challenge due to the complexity of its code. Moreover, sharing simulation setups among users is not straightforward, as it requires writing (C++) code in a specific, reusable manner. Several high-level software frameworks have been developed to address these issues and are currently used by a large community, such as GAMOS~\citep{arce2014}, TOPAS~\citep{faddegon2020} or GATE 9~\citep{sarrut2022}. All three target medical physics applications and rely on macro script commands (Geant4 macros for GATE and GAMOS, and an ad-hoc system for TOPAS). This approach allows for a simplified description of simulations, which is essential because users of such applications should focus on physics rather than the technical details. A disadvantage of macro commands is that they are limited to parameter descriptions with few or cumbersome instructions (loops, conditional statements, etc.) and are not easily extendable.

For GATE~10, we chose Python as the user language to configure and run simulations and as a user interface to the low-level Geant4 engine. The reason for this choice was that Python has become a major tool for data science over the past years, offering ease of learning and convenient access to powerful mathematical~\citep{harris2020} (NumPy), statistical, and artificial intelligence (AI) libraries written in C++ and optimized for high-performance computing (HPC). Instead of requiring users to learn a new syntax of commands specific to GATE, we recognized that a large proportion of users already know Python or will learn it during their career.

\subsection{Previous efforts to steer Geant4 via Python}
The Geant4 collaboration has previously proposed an initiative to steer Geant4 simulations via Python with the g4py or g4python\footnote{\url{https://github.com/koichi-murakami/g4python}} package, which to our knowledge is no longer developed or maintained. Although g4py provided a way to avoid C++ compilation of a Geant4 simulation, it still required users to have detailed knowledge of the Geant4 toolkit because it directly exposed low-level Geant4 classes and functions. Furthermore, g4py required complex installation procedures due to its dependence on the Boost library~\citep{schling2011}. Other solutions to expose Geant4 code to Python have also been proposed, including \verb|geant4_pybind|\footnote{\url{https://github.com/HaarigerHarald/geant4_pybind}}, but it is unclear how these projects will evolve. In any case, users must always compile Geant4 by themselves.
Additionally, GATE implements some of its own core functionality in C++, which needs to be exposed to Python.   
For these reasons, we designed a binding layer from scratch and included it into GATE~10. We explain the technical details of this new contribution in Part 2 (see companion paper~\citep{krah2025}).

\subsection{File Management and Output}\label{sec:file-management-output}

Simulations in GATE require a multitude of parameters and options, as well as access to external input data such as anatomical images, ROOT files, or material databases. We sought to achieve a balance between dependence on external software packages and the use of internal libraries.

Input images are handled by the ITK toolkit~\citep{mccormick2014}, which supports reading and writing a wide variety of image formats, including common types like MHD, NIfTI, and DICOM. ITK can manage images with different types of numerical data (e.g., float, int) and dimensions (2D, 3D, 4D). The reading and writing of images are performed by Python, leveraging ITK's Python wrapping. Once the images are loaded into memory, they are transferred to the C++ side and managed by the C++ part of ITK, ensuring more efficient and integrated image processing. Relying on ITK ensures robustness and access to many file formats. 

ROOT files are a standard file format for physics-related data, such as lists of particles (often referred to as ``trees'' or ``phase spaces''). This format is convenient for managing large file sizes and supports compression. Historically linked to the Geant4 community and developed at CERN, ROOT is familiar to many Geant4 users \citep{brun1997}. Moreover, as GATE has long used the ROOT output file format and GATE users are thus likewise familiar with it, we decided to keep it as the main output format for list-mode data. 

%

Additionally, GATE manages other types of data, such as material files that follow the traditional GATE format (i.e. text files describing material density, elements, etc.) or lists of energy spectra for conventional ICRP107 databases. In the future, we consider integrating file formats from the Emission Tomography Standardization Initiative (ETSI\footnote{\url{https://etsinitiative.org}}) once these are well established. ETSI is currently designing a completely new file format for preclinical and clinical emission tomography (PET and SPECT imaging) raw data, including list-mode data, singles and sinograms/histograms. This file format aims to facilitate more efficient data manipulation, storage and memory usage. In addition, having a common raw data file format for both simulated and measured data will make the workflow more efficient. For example, such integration will facilitate the smooth incorporation of GATE-generated output raw data into reconstruction software, enhancing the overall workflow and compatibility with industry standards.




\subsection{Transition from GATE~9 to 10}

The transition to GATE 10 requires users to convert their existing macro files into Python scripts and adapt to some changes in parameter names and logic. While the development aimed to preserve the feel of previous versions, we prioritized modifications that offered significant improvements. 

A key example is the replacement of the \verb+System+ concept for PET or SPECT detectors used to collect detected hits via a Geant4 Sensitive Detector: now, any volume can collect hits via an internal mechanism based on Geant4 Primitive Scorers.

We anticipate that users will require some time for transitioning from GATE~9 to GATE~10.
Therefore, the GATE collaboration has decided to continue maintaining the GATE~9 codebase, adapting it to new versions of Geant4 and providing minor bug fixes, but there will be no further development work on GATE~9. On the other hand, GATE~10 development will focus on consolidating the released code and on incorporating useful functionality that has been part of GATE~9 but not yet implemented in GATE~10. 

We have evaluated the feasibility of automatic tools to translate existing GATE~9 input files to GATE~10 but concluded that such tools would be unreasonably difficult to implement reliably. Instead, the online documentation will contain a growing section of examples of simulations that have been successfully translated by users. 


\subsection{Robust Development Cycle and Cross-Platform Compatibility}

Based on the experience gained from the open development of GATE, the challenge was to establish a robust framework to collaboratively develop new features and fix bugs while ensuring reliable validation, consistent results and timely updates aligned with Geant4 releases. The solution was to implement modern Continuous Integration (CI) practices and a comprehensive testing strategy. For every new feature added to the codebase, the following rules are enforced: (1) code must be submitted via the GitHub's Pull Request mechanism, (2) automated verification tests must be provided, and (3) documentation must accompany the submission.

Developing tests in a Monte Carlo context presents unique challenges. Firstly, the tests should ideally have short computation times, even though most simulations require long runtimes. Secondly, the outputs are stochastic, meaning comparisons with reference data must be performed statistically. Small variations in results are expected, which makes direct comparisons more difficult. This challenge has already been acknowledged by the Geant4 collaboration, which has developed specific physics-based tests \citep{arce2021, arce2025}.

Each feature to be tested is unique, so a set of global recommendations is adopted. The random engine generator seed must be fixed, and computation times should be kept "as low as reasonably achievable", with a maximum of 1-2 minutes. When comparisons with reference data are necessary, statistical tests such as Z-score, Wasserstein distance (for comparing distributions), or sum of squared differences (for images or dose maps) are employed. Each feature must have at least one associated test before being merged into the main branch. Each test ends with a single True/False result, indicating whether the test passed successfully. This process has proven effective; as of early 2025, more than 230 tests have been created.

Thanks to GitHub's CI infrastructure, every code modification automatically triggers the test suite. If any test fails, the code is not merged, and a discussion with the contributor is initiated to understand and resolve the issue. This approach enforces a minimal set of requirements for contributors: they must provide a formal test, and the CI system simplifies the process of verifying that new contributions do not break existing functionality. While tests are triggered automatically by CI, they can also be run locally on the user’s machine using the \verb|opengate_tests| command.

On GitHub, tests are conducted across multiple Python versions (from 3.9 to 3.12 at the time of writing) and various platforms (OSX, Linux, Windows), ensuring cross-platform compatibility. To reduce computational overhead, a Monte Carlo-inspired approach (pun intended) is used, where a randomly selected subset of tests is run on each platform for each commit. Additionally, a complete test suite is run once a week across all platforms and Python versions. Note that Windows compatibility is still not fully functional, and multithreading cannot yet be used together with ROOT output.

Finally, the core libraries are built and integrated into a single "wheel" (a Python package) that is uploaded to PyPI\footnote{\url{https://pypi.org/project/opengate/}}. This allows users to install GATE 10 with a single command: \verb|pip install opengate|. Under the hood, the process involves compiling several dependent libraries such as ITK and Geant4. Since Geant4 requires a large physical data library (approximately 2 GB), which cannot be included in the wheel due to size restrictions, a dedicated mechanism ensures that the correct data is automatically downloaded during installation if not already available. This mechanism also handles the automatic download of test data during the first execution of the \verb|opengate_tests| command. While the standard installation enables users to create simulations in Python, it does not include the C++ source code. For developers who wish to extend GATE 10 by adding new features, the installation of a specific developer option is essential, which involves compiling ITK and Geant4 from source.

In conclusion, the development cycle is managed through a test-based framework and continuous integration, ensuring reliable software quality and facilitating collaborative contributions. Everything is openly available on \url{https://github.com/OpenGATE/opengate} and the collaboration allows contributions from all over the world.

\section{New and enhanced features}

In this section, we describe features that have either significantly changed compared to the previous versions of GATE, or that have been newly developed and added to GATE~10. We group these into geometry, actors (scorers), physics, and sources.

\subsection{Geometry}

\subsubsection{Parallel worlds}

Parallel worlds in Geant4 provide a generic way to run simulations in different geometries simultaneously. A practical use case in medical physics arises when volumes overlap, such as the gantry of a linear accelerator or SPECT detectors with the patient’s CT bounding box or when the user defines an analytical volume inside a voxelized CT image (e.g., brachytherapy seeds). In GATE~10, users can use the Geant4 native parallel world logic with minimal additional configuration, as Geant4 mechanics are handled internally. However, a caveat of using parallel worlds is the increased tracking time, as the Geant4 engine tracks particles in all worlds, potentially doubling the computation time.

\begin{figure}
  \centering
  \includegraphics[width=0.45\columnwidth]{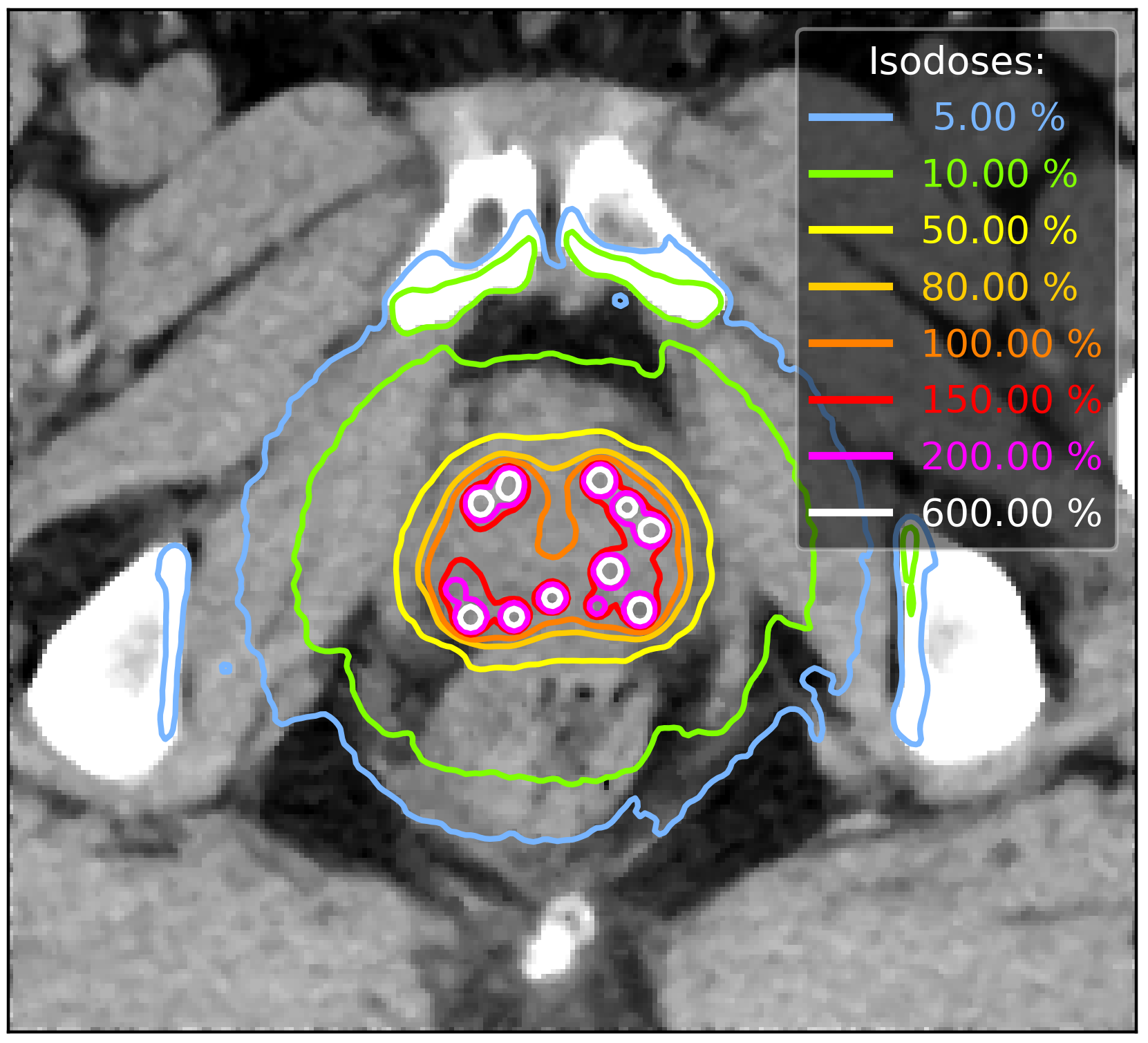}
  \caption{Example of the \textit{parallel worlds} feature through prostate brachytherapy treatment simulation. The isodoses, overlaid on the patient's CT scan, are normalized with respect to the prescription dose.}
  \label{fig:parallelworlds}
\end{figure}

A medical application that illustrates this feature is low-dose-rate prostate brachytherapy, where dozens of radioactive seeds (represented as analytical objects) are inserted into the prostate, as a voxelized CT volume. A dosimetry result, based on low-dose-rate clinical data, is illustrated in figure~\ref{fig:parallelworlds}. In this case, the computational phantom of the patient is defined in the ``main world'' while the 53 seeds defined in the treatment plan are described in a ``secondary (parallel) world''.

\subsubsection{Tessellated geometry}
\begin{figure}
  \centering
  \includegraphics[width=0.8\columnwidth]{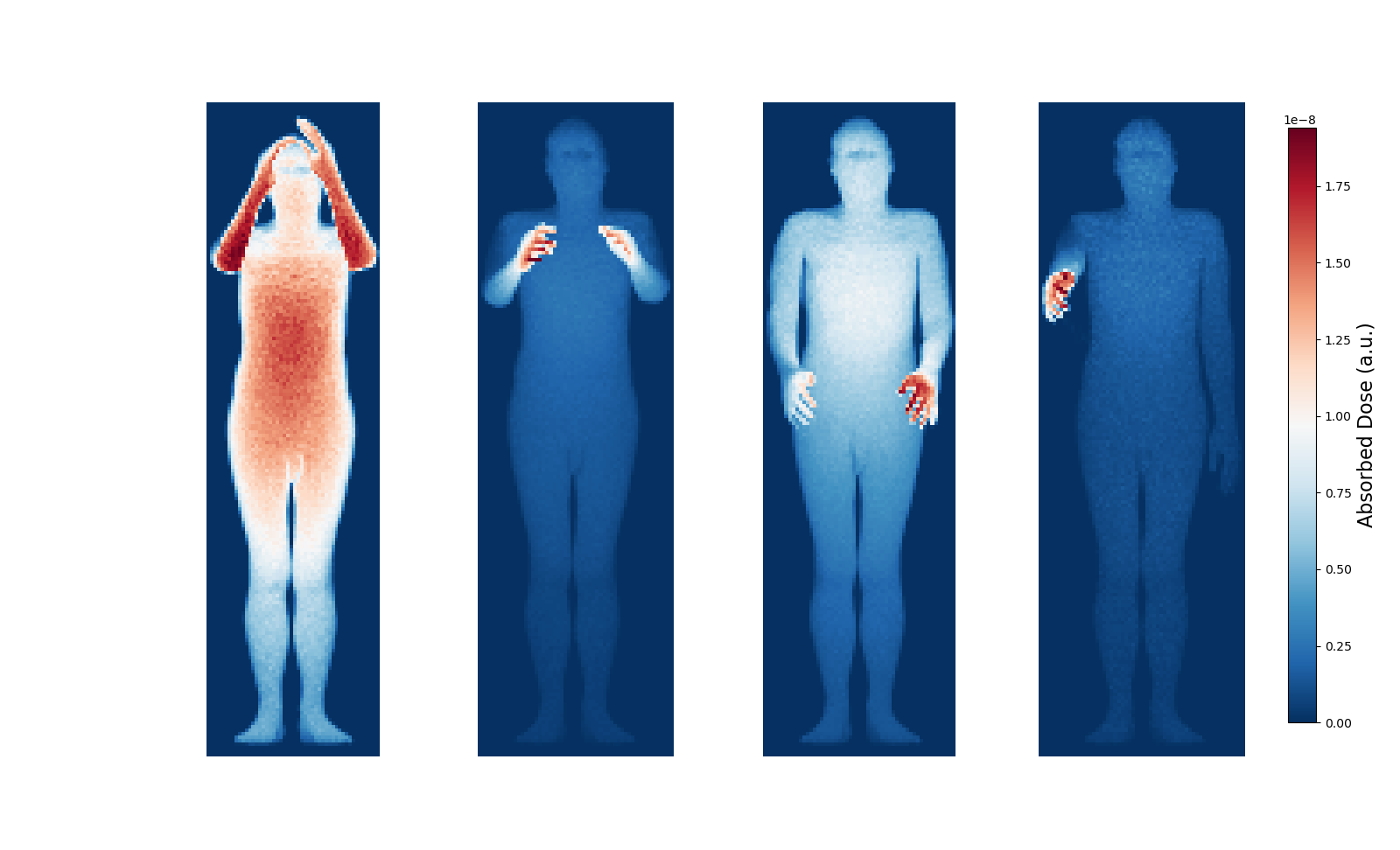}
  \caption{Example of the \textit{tessellated geometry} feature through skin dose exposure calculation in X-ray guided intervention, for different poses and positions of the physician around the patient.}
  \label{fig:STL}
\end{figure}

With the new tessellated geometry feature in GATE 10, the user can import mesh phantoms in stereolithography (STL) format and thus consider complex CAD models or fine-detail objects.
STL is a file format used to represent 3D models, describing the surface geometry of objects using a mesh of interconnected triangles. It is managed in Geant4 by the \verb=G4TessellatedSolid= class.
An example is the phantom provided by the ICRP~145 mesh dataset \citep{kim2020icrp}.
Figure~\ref{fig:STL} shows the GATE~10 simulation of a cardiologist's exposure to scattering radiation during an X-ray-guided intervention in the operating room. The simulations were based on different poses of the ICRP~145 phantom created using a third party software such as Blender and each pose was imported as an STL file. Another application is the implementation of hardware components which already exist as CAD models.

\subsubsection{Boolean solids}

Geant4 allows users to combine solids via boolean operations, i.e. to create intersections, unions, and subtractions of geometrically parameterised solids. Contrary to previous GATE versions, GATE~10 includes this functionality. Although the boolean operations are performed on Geant4 solids, we decided to implement boolean volumes, in analogy to other GATE volumes, and hide the underlying mechanisms from the user in favour of a simpler and more intuitive interface. The user needs to define and configure two volumes, e.g. ``A'' and ``B'', and then apply the boolean operation via a function, e.g. \verb|substract_volumes(A, B)|. The returned boolean volume is then included in the simulation representing the volume ``A-B''. Boolean solids are useful for constructing complex shapes such as complex collimators in SPECT, but at the expense of computing time as the navigation in a complex Boolean solid is proportional to the number of constituent volumes.

\subsubsection{Voxelized geometry}

In medical physics simulations, a CT image of a patient or a computational phantom is often used. It is represented as an array of voxels and requires special care to allow fast particle tracking in Geant4. From the simulation point of view, a voxel is a small box characterized by its material properties, through which particles need to be transported. In GATE, any three-dimensional input image can be used to create a voxelized geometry component (\verb|ImageVolume|) by providing a mapping from image values to materials as an input parameter (\verb|voxel_materials|). GATE does not make any assumptions about the nature of the input image, as long as it can be mapped to materials. Internally, this is managed by a fast Geant4 navigator based on nested parameterization using \verb=G4PVReplica= and \verb=G4PVParameterised= features~\citep{sarrut2008}. A common case in medical physics is the use a CT image in Hounsfield Units. To assist with this, GATE~10 provides a helper function, \verb|HounsfieldUnit_to_material|, which creates an image value-to-material lookup table.

\subsubsection{Voxelization of scene}

GATE~10 introduces a new concept by which the user can create a voxelized representation of the simulation geometry, i.e. of the Geant4 volumes such spheres, boxes, cylinders, etc. contained in a simulation. Through a single command (\verb|voxelize_geometry|), the user can transform the analytical described geometry into a discrete 3D grid of voxels, similar to an image. The user needs to specify the grid size (spacing) and can furthermore specify which volumes should be included in the voxelized representation, e.g. by considering only a certain subset of volumes such as an imaging sensor. The function \verb|voxelize_geometry()| returns an ITK image, which can be saved to disk, and a dictionary that maps voxel labels to the materials of the geometry scene. \verb|voxelize_geometry()| performs the conversion by iteratively querying the Geant4 engine for the material at each voxel's center to determine the voxel label.

\subsection{Actors and scorers}

\subsubsection{Phase space actor}

The phase space actor in GATE records detailed information about particles reaching a specified volume during a simulation. By attaching this actor to a specific volume, users can obtain comprehensive insight into particle interactions with the volume boundaries. By default, the phase space actor records information about particles entering the volume, but it can be configured to store data about particles exiting the volume or particles undergoing interactions in a volume for the first time.
The recorded phase space data can be used as an input source for subsequent simulations, enhancing workflow efficiency.

In GATE~10, more than 50 attributes are currently available for storage, including key parameters related to particle history, such as parent particle information, volume of generation, position, direction, energy, time, and track length. Users have the flexibility to select the attributes to be saved and can thus optimize data storage and read/write speed. The phase space actor generates output in the ROOT file format, which can be easily analyzed using Python libraries such as uproot~\citep{jim_pivarski_scikit-hepuproot_2020}.

\subsubsection{Digitizer actors}

Introduced in an earlier version of GATE~\citep{jan2004,kochebina2024}, the Digitizer is a key component of the GATE workflow for PET, SPECT, and other imaging applications. It simulates signal readout and electronic distortions using analytical and semi-analytical models. It reproduces effects on energy, position, and time resolutions as well as transport and detection efficiencies. This results in a realistic simulation of the entire signal processing chain. GATE provides around ten digitizer actors, that can be arranged by a user in a processing chain to reproduce the required effects. ROOT output can be saved at each step of this digitizer chain.

GATE~10 also includes a Coincidence Sorter for PET or Compton Camera applications. It pairs single events (singles) that occur within a user-defined coincidence time window. Coincidences with oblique lines of response can be excluded by limiting the axial distance between their two singles. Likewise, coincidences between adjacent detectors can be excluded by imposing a minimum transaxial distance. If multiple coincidences fall within the same window, the user can choose from six available policies to resolve these cases (e.g., select the most energetic event or ignore multiple events). Currently, the Coincidence  Sorter operates in an ``offline'' mode as a Python script, which can be executed after the simulation to process the output list of singles. This functionality allows users to save and reload intermediate digitizer output, enabling efficient offline coincidence sorting.

\subsubsection{Enhanced linear energy transfer (LET) actor}

In GATE~10, both track- and dose-averaged LET (unrestricted) can be calculated. By default, it employs method C from \citet{cortes-giraldo2015}. Currently, stopping powers are obtained using \verb|G4EmCalculator|, but future updates aim to support user-provided custom stopping power tables. The output of this scorer is a 3D matrix of LET values.

\subsubsection{Relative biological effectiveness actor}

The \verb|RBEActor| in GATE 10 enables the calculation of relative biological effectiveness (RBE) and RBE-weighted dose based on the local effect model I (\verb|LEM1lda|)~\citep{kramer_rapid_2006} and a modified microdosimetric kinetic model (\verb|mMKM|)~\citep{inaniwa_treatment_2010}. The actor supports multithreading and multiprocessing in GATE 10. Users provide lookup tables for the intrinsic parameter ($\alpha_z$ for \verb|LEM1lda|) or dose-averaged saturation-corrected specific energy ($z^*_{1D}$ for \verb|mMKM|) within the desired proton/ion energy ranges, which are used for scoring per simulation step. Additionally, users can customize cell radiosensitivity parameters and model specific parameters (\verb|D_cut| and \verb|r_nucleus| for \verb|LEM1lda|, \verb|alpha_0| and \verb|F_clin| for \verb|mMKM|), allowing flexibility in radiobiological modelling.


\subsubsection{TLE (hybrid) actor}

The Track Length Estimator (TLE)~\citep{baldacci_track_2014} is a Variance Reduction Technique (VRT) designed to rapidly compute the deposited energy and dose in voxelized geometries for low-energy gamma rays, typically below 1-2 MeV, where electron delocalization is minimal. This method is particularly effective in scenarios where precise energy deposition calculations are needed without the computational overhead of detailed particle tracking. In GATE 10, we have adapted the TLE implementation to include a hybrid mode that automatically switches between tracking gamma dose with and without the TLE method, based on the gamma rays energy. This hybrid approach optimizes the simulation process using the TLE method for lower energy gammas, where it is most efficient, and switching to standard tracking for higher energy gammas. Users can easily activate this feature by declaring the TLE actor attached to a specific volume. The TLE method has been previously validated and has demonstrated significant speed improvements (up to 100 times faster) ~\citep{noblet_validation_2016} depending on the configuration. Note that the enhanced versions seTLE (Split Exponential)~\citep{smekens_split_2014} and neutron~\citep{elazhar2018} is not yet implemented in GATE 10.

\subsubsection{Free flight actor}

A new (approximate) variance reduction technique has been introduced, particularly beneficial for Monte Carlo simulations of SPECT. This method introduces the concepts of Free Flight, a biasing technique from Geant4, and Angular Acceptance rejection in a two-step approach, addressing both primary and scattered events. Primary gamma photons are tracked in straight lines towards the detector, considering angular acceptance and weighted contributions. Scattered events are simulated using a combination of split Compton and Rayleigh interactions, which are then free-flighted towards the detector. When compared to standard analog Monte Carlo simulations, this technique demonstrated a speedup of approximately 50-fold, with even greater improvements in low-count regions. A significant advantage of this method is its flexibility: it can be applied to various geometrical elements in the simulation scene without requiring a detector response function, such as an Angular Response Function (ARF). This makes it especially suitable for the development of SPECT collimator designs.

\subsection{Sources and physics}

\subsubsection{Physics lists and physics options}

GATE~10 provides a transparent interface to the extensive library of physics lists available in Geant4. Users can easily enable various validated physics models via simple commands, without any modification to the underlying Geant4 physics engines or databases. This approach ensures consistency with the Geant4 toolkit and allows users to leverage its continuous improvements. Available physics lists include, but are not limited to: Electromagnetic Physics (Standard, Low Energy e.g. Livermore, Penelope), Hadronic Physics (FTFP\_BERT, QGSP\_BIC, QBBC, etc.), Radioactive Decay.

Beyond selecting these lists, specific processes and models can be enabled. For example, the quantum entanglement of annihilation photons, introduced in Geant4~11.0, can be activated using a dedicated command (\verb|g4_em_parameters.SetQuantumEntanglement(True)|). The standard two-gamma annihilation process is naturally included in these lists; however, triple-gamma annihilation is not yet available. Atomic de-excitation processes such as fluorescence, the Auger cascade, and Particle-Induced X-ray Emission (PIXE) can be activated with the Python interface to the Geant4 \verb+G4EmParameters+.

Positronium Lifetime Imaging (PLI) is a relatively recent multiphoton imaging technique that extends conventional PET tomography by incorporating additional information related to positronium—a metastable bound state of an electron and a positron~\citep{shibuya2020,moskal2021, Qi2022, tashima2024,steinbergerPositroniumLifetimeValidation2024,  moskal2024,Huang_2024, Mercolli2025.05.28.25328504}. Positronium lifetime distribution can provide complementary diagnostic insights. To measure positronium lifetime, non-pure $\beta^{+}$ emitters that also emit prompt gamma photons are required.
Accurate simulation of positronium decays demands the inclusion of various decay channels and careful modeling of the lifetime component mixtures within phantoms. In GATE (starting from v9.3), a set of helper classes has been introduced to support positronium decay simulations, enabling partial modeling of positronium-related physics. The implemented decay models include: para-positronium (two-photon decay), ortho-positronium (three-photon decay), and a mixed model where users can define the relative proportions of each decay type. All relevant parameters can be configured using standard GATE macros. Additionally, the emission of a prompt gamma with a user-defined energy can be simulated.
Ongoing work aims to port this functionality to GATE 10 and extend it to support more comprehensive positronium physics simulations, particularly to facilitate the modeling of complex phantoms.



\subsubsection{Annihilation Photon Acollinearity (APA)}

The spatial blurring in positron emission tomography (PET) images caused by the annihilation photon acollinearity (APA) becomes more pronounced as the effective diameter of the scanner increases. 
This effect is known to follow a Gaussian distribution. Toussaint et al. showed~\citep{toussaint2024} that APA was incorrectly modeled in various Monte Carlo software, which resulted in an underestimation of its effect on spatial resolution. 
Their corrections and conclusions were included in GATE 10.

In GATE, the user can define a source for PET systems in 3 different ways: as an ion source (e.g. of \isotope[18]{F} or \isotope[68]{Ga}), as a positron source (with the corresponding energy spectra), or as a back-to-back gamma source (neglecting positron travel). By default, annihilation photon pairs from positron-electron annihilation will be collinear. However, it is known that, for example for water between 20–30°C, the deviation of APA follows a 2D Gaussian distribution with a FWHM of 0.5°~\citep{colombino1965}. For an ion or an electron source, to enable this behavior, the user should set the \verb=MeanEnergyPerIonPair= of all the materials where APA is desired to 0.5 eV. The property is defined for a given list of materials, not for a specific volume. In other words, if a simulation contains two water volumes, one where annihilation photons have acollinearity and one where they do not, two variants of ``water'' need to be defined. For a back-to-back source, activation of APA is enabled by a boolean flag and the FWHM of its 2D Gaussian distribution can be defined by the user (with 0.5° being the default). The implementation of APA for back-to-back sources is based on assuming that its deviation follows a 2D Gaussian distribution which is a simplification of the true physical process.

Shibuya et al. \citep{shibuya2007} have shown that the deviation of APA in a human subject follows a double Gaussian distribution with a combined FWHM of 0.55°. While the double Gaussian distribution is currently not available in GATE, setting the \verb=MeanEnergyPerIonPair= of the material to 6.0 eV results in a 2D Gaussian with a FWHM of 0.55\degree.

\subsubsection{Scanned pencil beam sources}

Like previous GATE versions~\citep{grevillot2011monte}, GATE 10 enables users to simulate the delivery of scanned ion pencil beam treatment plans, using the \verb|IonPencilBeamSource| and \verb|TreatmentPlanPencilBeamSource|. In this treatment modality, the target is irradiated by scanning the beam across the transverse plane with steering magnets, while adjusting the beam energy to reach different depths within the target.
While retaining all the features existing in GATE 9, the new version of the source introduces the capability to directly read treatment plan files in their native DICOM format, thereby eliminating the need for prior conversion of the irradiated spot information into a text file. Additionally, it offers greater flexibility by allowing users to configure the source without a treatment plan, using custom spot data instead.

\subsubsection{Voxelized sources}

An activity source can be transformed into a voxelized source, using the \verb|voxelized_source| function, providing the labels-volumes-activity mapping. An image is then generated, in which each voxel is assigned to an activity value. A 3D activity distribution from an image can be used as a voxelized source, by creating a \verb|VoxelSource| object setting the input image for the corresponding parameter. Several other parameters of the source object should be set accordingly (particle type, activity, etc.). Particles will then be randomly located in the image voxels according to the given 3D probability distribution. For example, a SPECT image can be used as activity map and the voxelized CT scan of the patient as attenuation map (assigned to the \verb|attached_to| parameter). A useful function is provided in GATE 10 which aligns the voxelized activity with the CT image by computing the correct translation (\verb|get_translation_between_images_center()|).

\subsubsection{TAC sources}

The time-activity curve (TAC) feature can now be enabled for a particle source. The TAC is represented as a discrete histogram parameterized by two vectors: time points and corresponding activities. When enabled, the source's activity during the simulation is updated based on the current simulation time using linear interpolation of the TAC. If the simulation time is earlier than the first time point or later than the last one in the time vector, the activity is set to zero. The number of elements in the time vector determines the accuracy of the TAC representation. This feature is useful for simulating dynamic processes where the activity of the source changes over time, due to radioactive decay and tracer kinetics.

\subsubsection{PHID source}

A novel virtual source model named PHID (Photon from Ion Decay) was designed to simulate photons emitted during the complex decay chains of alpha-emitters, making it useful for SPECT acquisition simulations~\citep{sarrut2024}. The model extracts photon emission lines for both isomeric transition and atomic relaxation processes from the Geant4 database for a given alpha-emitter. The Bateman equations are used to calculate photon abundances and activities throughout the decay chain, considering decay rates and initial radionuclide abundances over a specified time range. PHID generates photons with accurate energy and temporal distributions without simulating the entire decay chain, resulting in a speed-up of the simulation. For example, in a simulation of 1 MBq of \isotope[225]{Ac} in water, PHID demonstrated a 30$\times$ speedup compared to analog Monte Carlo simulations. Additionally, compared to a simplified source model using only the two main photon emission lines, PHID simulated twice as many particles and detected 60\% more counts in the resulting images.

\subsubsection{GAN sources}

Particle tracking within a voxelized patient may be computationally inefficient because each voxel intersection triggers a new particle step. Recently, the development of a new approach combining low-statistics Monte Carlo simulations with AI was proposed. The outgoing gamma photons from a patient's body are recorded at the body surface and used to train a deep neural network based on a Generative Adversarial Network (GAN~\citep{goodfellow2014}). Once trained, the GAN’s generator can replicate the probability distributions of these particles, including their position, direction, and energy, effectively serving as a low-cost particle source. With conditional GANs (cGANs), a single low-statistics simulation is sufficient to train an activity-parameterized family of GANs, enabling the generation of particles for any desired activity distribution. Coupled with Angular Response Functions (ARFs), this technique has shown speed-ups of more than 100 times in image generation~\citep{sarrut2021b, saporta2022}. However, the approach requires specific cGAN training for each CT image and customized ARF models for different detector and energy window configurations. GAN sources can also be useful for modeling linear accelerator (linac) phase-space data~\citep{sarrut2019a, sarrut2021c}.
All these approaches have been integrated into GATE~10, facilitated by the Python interface. These methods rely on PyTorch~\citep{paszke2019}, a widely used deep learning framework, which must be installed to enable their functionality.

\subsubsection{Optical Simulation using GANs: optiGAN}

Optical photon transport simulations are essential for developing and optimizing radiation detectors in medical imaging and high-energy physics. While detailed optical Monte Carlo methods remain the gold standard for modelling photon interactions, their high computational requirements pose a challenge, especially when considering large system simulations. To address this, a Generative Adversarial Network (GAN) model, optiGAN, was developed \citep{trigila_generative_2023}, optimized \citep{srikanth_gpu_2024}, and integrated into GATE 10 \citep{mummaneni_deep_2025}. OptiGAN reduces the optical photon transport simulations time to half while preserving modelling accuracy with more than 92\% similarity between traditional and GAN simulations in GATE 10 \citep{trigila_generative_2023,mummaneni_deep_2025}.

An optical simulation using optiGAN provides very similar results as the full optical simulation but without the need of optical photon tracking. Users must define both the geometry and source. Currently, GATE 10 includes a specific model checkpoint from offline training on a predefined dataset (a $3 \times 3 \times 10$ mm$^3$ bismuth germanate crystal). This configuration is readily available for testing, with examples provided on GATE's GitHub. However, since the implementation is open-source~\citep{roncalilabgithub}, users will be able to generate their own datasets to train AI models, and integrate them into GATE 10.

\subsection{Automatic serialization into JSON file}\label{sec:serialization}

GATE~10 includes an automated feature which writes the configuration of the entire simulation into a JSON file, i.e. a structured, human-readable file. JSON is a widely adopted data format supported by a multitude of libraries. The user can activate this feature by setting \verb|sim.store_json_archive = True|. The manager-based architecture, together with functionality in the base class \verb|GateObject| (see the second part of this paper~\citep{krah2025}), automatically collects all relevant user input parameters of all components of a simulation and serializes them into JSON format.
The generated JSON file can be reloaded into GATE~10, and the simulation will be recreated based on it. 
This feature is particularly useful when a user runs multiple simulations from a single Python script, e.g. with different geometry configurations, and wishes to maintain access to parameters of each of them. It is also helpful as part of an archiving mechanism when GATE 10 is used for automated routine dose calculation tasks, such that any simulation can be re-run later, although it was generated dynamically by a script.

\subsection{Dynamic parameterization, timing}\label{sec:dynamic-param}

A Geant4 simulation, by itself, is only time-aware with respect to tracking of primary particles. However, many simulations in medical physics and imaging require time-dependent modeling on a macroscopic time scale. For example, the source rotates around the patient, the patient breathes, the activity distribution changes due to kinetics, etc. To accurately reflect such dynamics, simulations must be discretized in time. A brute force approach would involve running a simulation multiple times, each time, e.g., with a different gantry angle. Geant4 provides the concept of \emph{runs}: the simulation is initialized once at the beginning, and certain aspects of the simulation can be changed after all primary particles which belong to one run are tracked. Previous GATE versions implemented Geant4's concept of runs in a task-specific fashion. 
In GATE~10, we adopted a more generic approach through ``dynamic parameterisation''. Users can now define a sequence of run timing intervals for a GATE simulation, i.e. pairs of moments in time when a run is considered to start and stop, respectively. On the other hand, the user can provide a list of parameters to a component of a simulation instead of only providing a single parameter. For example, if a simulation is split into 8 run intervals, the user might provide a list of 8 CT images via the command \verb|add_dynamic_parametrisation()| to dynamically parameterise the patient geometry (e.g. to account for breathing). Additionally, the user might provide a list of 8 rotation matrices to rotate the particle source around the patient. The generic ``dynamic parameterisation'' approach provides a unified user interface, yet it is flexible and can be easily extended to other compatible parameters in the future.

\section{Discussion}

In this first article, we have presented GATE version 10, detailing its new main features and the principles guiding its development. GATE has transitioned from a standalone application to a versatile software platform that can be used not only for conventional simulations but also as a library integrated into other software packages. This latter capability is particularly important as it will allow GATE to be embedded into dedicated software, such as treatment planning systems or image reconstruction frameworks.

We outlined the impact of such highly collaborative scientific open-source software as a significant contributor in advancing the use of physics in medicine, biology and beyond. This impact can be evidenced by the publication of thousands of articles and citations of the four main GATE publications. Furthermore, several major companies utilize GATE for designing and optimizing their imaging or radiation-based prototypes and commercial systems. Additionally, GATE is used as a valuable resource in teaching several medical physics courses by providing a range of well-prepared computational laboratory materials.

We also highlighted the collaborative and community aspects, both interests and challenges, of the GATE initiative. The collaboration is open, governed only by a contribution-based model, allowing any group to integrate as soon as they propose a relevant and useful contribution. However, over time, some groups may stop contributing to the further development of GATE, necessitating a clearer definition of our vision and how resources should be maintained.

Due to its significantly lower learning curve compared to the underlying Geant4 toolkit, we think that GATE with Python can be a key educational tool. It allows students to focus on the principles of physics rather than the complex programming details. As an example, students from the French DQPRM qualification (``Diplôme de Qualification en Physique Radiologique et Médicale'') follow Monte Carlo courses using GATE. We believe that this new version will further support and enhance the education of medical physicists in Monte Carlo simulation techniques.

Despite its advancements, GATE 10 remains an ongoing development with several limitations. First, there is a deliberate break in backward compatibility, which requires users to manually transcribe legacy macros into Python scripts. Technical limitations also exist; full functionality on the Windows platform is not yet achieved, and multithreading is currently incompatible with ROOT output. Furthermore, not all of Geant4's fine-tuning capabilities are exposed through the Python interface, potentially requiring advanced users to implement custom C++ code for specific needs. Some data processing modules are also still being finalized; for instance, coincidence sorting currently operates as an offline script, and digitizers for time-related effects like dead time and pile-up are in development.

Looking ahead, the new architecture opens up several avenues for future development. One area is the extension of GATE's capabilities into radiobiology, with better integration/link with the Geant4-DNA toolkit~\citep{incerti_geant4-dna_2010, arce_results_2025} for track structure simulations at the cellular and subcellular level. We also plan to expand the portfolio of variance reduction techniques ; for example, the free-flight method could be adapted to estimate out-of-field dose distributions in radiation therapy. Finally, while the Python interface already facilitates the use of neural networks, future work will focus on a tighter integration of AI-based models.

\section{Conclusion}

Looking beyond GATE 10's specific features, it is crucial to acknowledge the broader context in which such scientific software exists and evolves. The current development model was as follows: GATE is developed feature by feature by researchers for researchers, typically during their PhD or postdoc periods, supervised by their advisors. However, only few full-time researchers or engineers can devote time to coding. Funding for such human power is crucial, but has, until now, been associated with research grants as a side product of conventional research publications. This makes it challenging to sustain a continuous development, maintenance, and support of such a large and complex program challenging and slow, as specific funds for these service tasks have been a rare exception.

As we look into the future, the interest of the OpenGATE collaboration is to continue to innovate and adapt GATE according to the significant advances of computational and software methods to meet the evolving needs of medical physics and beyond. The challenges ahead will require sustained large-scale collaboration, a continuous stream of funding, and a commitment to maintaining and enhancing this valuable toolkit. We hope that GATE will continue to serve as a cornerstone in the advancement of biomedical imaging and radiation therapy, fostering a community of researchers and developers dedicated to pushing the boundaries of scientific discovery and technological innovation.

In the companion paper~\citep{krah2025}, we detail technical challenges and selected architectural innovations.

\section*{Acknowledgments}

The authors acknowledge funding support from the following grants: ANR-21-CE45-0026, INCa-INSERM-DGOS-12563, ANR-11-LABX-0063, ANR-11-IDEX-0007, Erasmus+ program. 


\bibliography{bibliography.bib}

\end{document}